\documentclass[conference]{IEEEtran}
\usepackage{cite}
\usepackage{amsmath,amssymb,amsfonts}
\usepackage{algorithmic}
\usepackage{graphicx}
\usepackage{textcomp}
\usepackage{xcolor}
\usepackage{tabularx}  
\usepackage{colortbl}  
\usepackage{booktabs}  
\usepackage{makecell}  
\usepackage{multirow}  
\usepackage{enumitem}  
\usepackage{adjustbox}  
\usepackage{relsize}
\usepackage{hyperref}

\usepackage{balance}

\usepackage{mdframed}
\usepackage{lipsum}

\makeatletter
\def\ps@IEEEtitlepagestyle{%
  \def\@oddfoot{\mycopyrightnotice}%
  \def\@evenfoot{}%
}
\def\mycopyrightnotice{%
  {\footnotesize
  \begin{minipage}{\textwidth}
    \textcopyright 2019 IEEE.  Personal use of this material is permitted. Permission from IEEE must be obtained for all other uses, in any current or future media, including reprinting/republishing this material for advertising or promotional purposes, creating new collective works, for resale or redistribution to servers or lists, or reuse of any copyrighted component of this work in other works.
  \end{minipage}\hfill}
  \gdef\mycopyrightnotice{}
}

\newmdtheoremenv{Hypo}{Hypothesis}

\def\BibTeX{{\rm B\kern-.05em{\sc i\kern-.025em b}\kern-.08em
    T\kern-.1667em\lower.7ex\hbox{E}\kern-.125emX}}

\begin{document}
\title{
    Renovating Requirements Engineering\\
    \smaller\smaller
    First Thoughts to Shape Requirements Engineering as a Profession
}

\author{%
    \IEEEauthorblockN{Yen Dieu Pham, Lloyd Montgomery, Walid Maalej}
    \IEEEauthorblockA{
    University of Hamburg \\
    Hamburg, Germany \\
    Email: \{pham, montgomery, maalej\}@informatik.uni-hamburg.de
    } 
}
\maketitle

\begin{abstract}
Legacy software systems typically include vital data for organizations that use them and
should thus to be regularly maintained.
Ideally, organizations should rely on Requirements Engineers to understand and manage changes of stakeholder needs and system constraints. 
However, due to time and cost pressure, and with a heavy focus on implementation, organizations often choose to forgo Requirements Engineers and rather focus on ad-hoc bug fixing and maintenance. 
This position paper discusses what Requirements Engineers could possibly learn from other similar roles to become crucial for the evolution of legacy systems.
Particularly, we compare the roles of Requirements Engineers (according to IREB), Building Architects (according to the German regulations), and Product Owners (according to ``The Scrum-Guide").
We discuss overlaps along four dimensions: liability, self-portrayal, core activities, and artifacts. 
Finally we draw insights from these related fields to foster the concept of a Requirements Engineer as a distinguished profession. 
\end{abstract}


\section{Introduction}

Architecture\footnote{In this paper,  with the term ``Architecture" we refer to building architecture.} related activities can be dated back to the ancient Greeks, Egyptians and Romans.
However, ``Architect" was first recognized as a profession around the 16th century.
Today, the term ``Architect"  means a qualified, educated, and indispensable professional for any kind of building projects \cite{WEB_Arch19}.
The services of Architects differ between countries \cite{WEB_FAGA11}.
German Architects, for instance,  provide design services as well as construction management, bidding, and cost controlling.
American Architects, however, will often only conduct the design \cite{WEB_FAGA11}. 
This broad range of responsibilities have led German Architects to specify legally binding activities to "protect the profession" and assure a high quality outcome for new construction and renovation projects.

Fowler has argued that software and buildings are incomparable because it is far more difficult to make changes to an existing building than to software \cite{Fow03}.
And yet many authors including Bisbal et al.~argued that it is also expensive to maintain legacy systems and ``difficult, if not impossible, to extend" them \cite{BLWG99}.
We argue that there are insights that can be drawn from the profession of Architects for Requirements Engineers, particularly in terms of maintaining and re-engineering legacy systems. 
This constitutes the main motivation of this work.

Similar to existing buildings, legacy software systems must be maintained and re-engineered to remain efficient and to satisfy changing stakeholder needs and system usage contexts  \cite{BCMV03}.
Ideally, Requirements Engineers should monitor and regularly analyze current stakeholder needs and constraints, to then recognize requirements changes and initiate system updates. 

Agile practices like Scrum should enable organizations to  efficiently react to changes \cite{WEB_AA19}.
Yet agile-oriented organizations often focus on implementation and ``ad-hoc maintenance'' and forgo Requirements Engineering practices \cite{PGU14}. 
An agile team would typically ``collect'' and translate requirements into tasks or action items in backlogs. But, little is known about where do these requirements come from, how they are collected, by whom, and whether they comply to previous decisions and technological, economic, or sustainability goals. 
Originally designed for small, collocated, knowledgeable, and empowered teams, agile practices are nowadays very popular among practitioners for every context \cite{Bas13} -- even if less suited, e.g. for projects targeting large or legacy systems \cite{SIJ14}.

This paper investigates and compares the roles of Requirements Engineers, Product Owners (found in agile organizations), and Architects. Our goal is to get insights and derive hypotheses on how to foster Requirements Engineers as  indispensable role in software projects, in particular concerning legacy systems. Each of these three roles is outlined in a set of guidelines or legal documents. We thus use these documents as basis for our comparison. 

For Requirements Engineers we relied on two documents that represent a knowledge baseline to become a certified professional for Requirements Engineering  \cite{Poh16}: the International Requirements Engineering Board (IREB) Study Guide ``Requirements Engineering Fundamentals" \cite{Poh16}, and the ``CPRE Foundation Level - Syllabus" \cite{WEB_CPRE17}.
IREB\footnote{\url{www.ireb.org}} is a non-profit organization, which consists of leading RE representatives from industry and academia.
For Product Owners we used ``The Scrum Guide"\cite{WEB_SS17}.
We have chosen this agile framework due to its popularity in industry and academia \cite{SH16}.

For Architects we used two documents: the ``HOAI - Fee Regulations for Services by Architects and Engineers" \cite{Spr13}, and ``Neufert - Architects Data" \cite{NNK12}.
HOAI sets out general provisions and definitions for architectural and engineering services, and Neufert is a standard reference for the initial design and planning of building projects.
We compare the documents against the statements of the ``The Federal Chamber of German Architects" \cite{WEB_Fed19}.

\section{Comparison Results}

\begin{figure}[!b]
    \centering
    \includegraphics[width=\columnwidth]{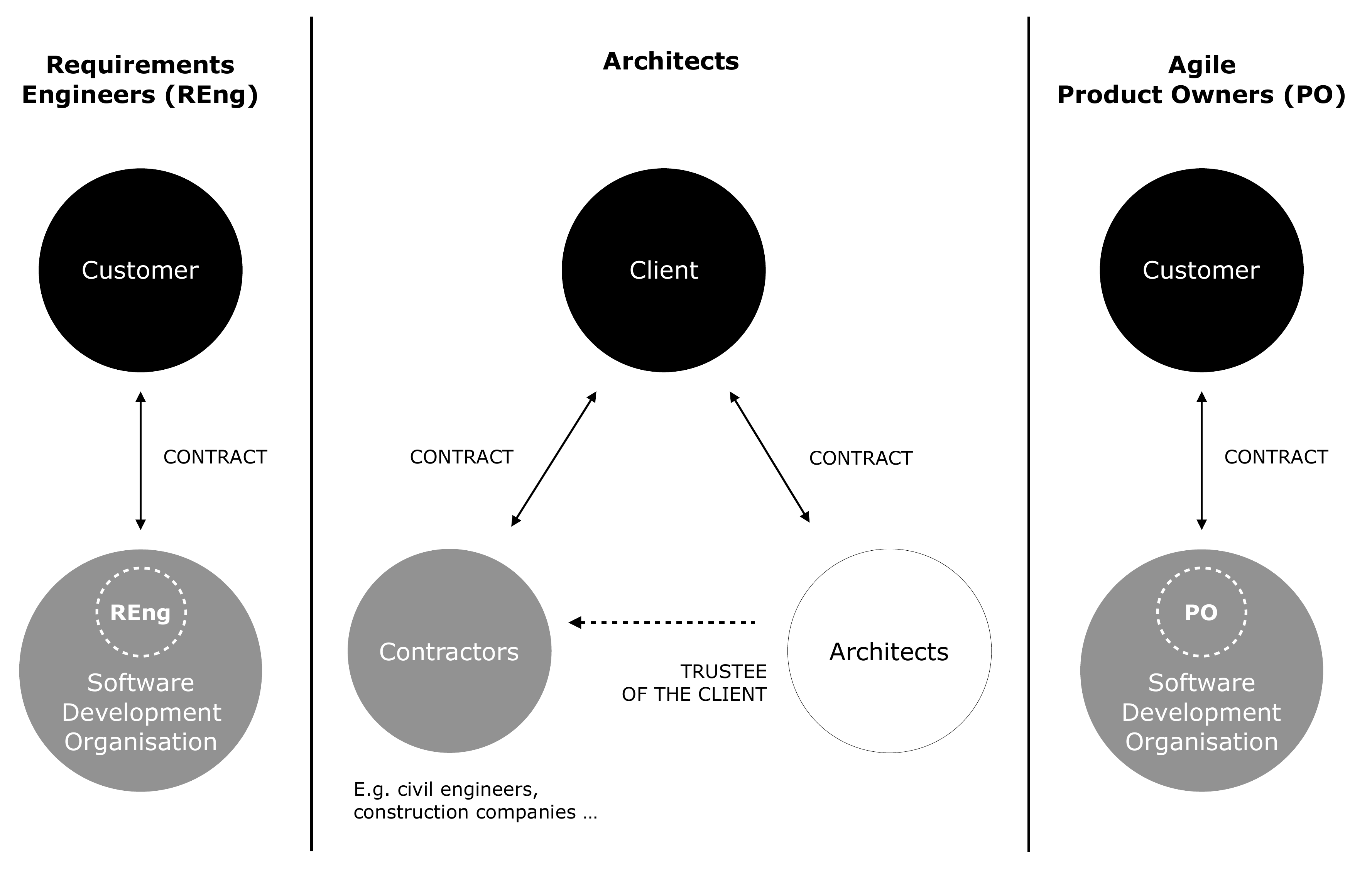}
    \caption{Liability of Architects vs. Requirements Engineers.}
    \label{fig:liability}
\end{figure}

Through an informal analysis, four core dimensions have emerged: ``liability", ``self-portrayal", ``core activities", and ``artifacts".
We further divided each of these core areas to help understand the extracted comparative elements.
With regard to self-portrayal, we extract the highlighted values and characteristic.
We match the “core activities” to the categories defined by IREB and describe the way the artifacts are implemented and used by each role or profession respectively.
Figure \ref{fig:liability} shows the \textit{liability} findings.
Table \ref{table:results} shows the  findings along the other dimensions. 
We discuss similarities and differences along the dimensions to reflect on current practices of the roles. 

\newcommand{\headerTop}[1]{\multicolumn{1}{c}{\textbf{#1}}}
\newcommand{\headerMiddle}[1]{\multicolumn{4}{c}{\larger #1}}
\newcommand{\colTitle}[2]{%
    \parbox[t]{2mm}{\rotatebox[origin=c]{90}{\textit{\textbf{\makecell{%
        #2 %
    }}} }}\vspace{#1mm}%
}
\newcommand{\colContent}[1]{%
    \begin{tabular}[t]{%
        @{\textbullet~}p{.95\linewidth}@{}}#1%
    \end{tabular}%
}
\begin{table*}[p]
\centering
\caption{Results of Comparison between Requirements Engineers, Architects, and  Product Owners}
\label{table:results}
\begin{tabular}{ m{.03\textwidth} | p{0.29\textwidth} | p{0.29\textwidth} | p{0.29\textwidth }}
    \toprule
    \headerTop{Title} &
    \headerTop{Requirements Engineers} & \headerTop{Architects} & \headerTop{Product Owners} \\

    \midrule
    \headerMiddle{Self-Portrayal} \\
    \midrule
    \colTitle{-16}{Highlighted values\\and characteristics}
    &
    \colContent{
        Analytical thinking
        \\Empathy
        \\Communication skill
        \\Conflict resolution skills
        \\Moderation skills
        \\Self-confidence
        \\Persuasiveness
    } & \colContent{
        Trustee of the client
        \\Holistic
        \\Persuasive regarding innovative design and construction
    } & \colContent{
        Commitment
        \\ Courage
        \\ Focus
        \\ Openness
        \\ Respect
    } \\

    \midrule
    \headerMiddle{Core Activities} \\
    \midrule
    \colTitle{-3}{Elici-\\tation}
    & \colContent{
        Obtaining requirements from stakeholders and other sources
        \\ Refining requirements
    } & \colContent{
        Clarifying tasks and dependencies
        \\ Explaining tasks and dependencies
    } & \colContent{
        Refining the items in the Product Backlog
        \\ Collaborating with the development team for refinement
    } \\\hline

    \colTitle{-8}{Documen-\\tation}
    & \colContent{
        Describing elicited requirements adequately 
    } & \colContent{
        Creating artifacts based on type, size, and progress of the project to the required extent and level of detail 
        \\ Considering all disciplines and individual-specific requirements
    } & \colContent{
        Clearly expressing Product Backlog items
    } \\\hline

    \colTitle{-10}{Validation and\\negotiation}
    & \colContent{
        Involving correct stakeholders
        \\ Separating identification and correction of errors
        \\ Repeating validation
        \\ Validating from different views
        \\ Constructing development artifacts
        \\ Changing documentation type adequately
    } & \colContent{
        Coordinating with involved parties
        \\ Inspecting artifacts and the contribution of all involved parties
        \\ Consulting the client
        \\ Negotiating with involved parties
        \\ Planning the construction
    } & \colContent{
        Inviting Attendees including Scrum Team and key stakeholders to Sprint Review
        \\ Ensuring the development team understands items in the Product Backlog to the level needed
        \\ Explaining what Product Backlog items have been ``Done" and what has not been ``Done" in Sprint Review
    } \\\hline
    
    \colTitle{-10}{Management}    
    & \colContent{
        Assigning attributes
        \\ Defining views on requirements
        \\ Prioritising requirements
        \\ Tracing requirements
        \\ Versioning requirements
        \\ Managing requirements changes
        \\ Measuring requirements
    } & \colContent{
        Adapting artifacts based on progress and contributions
        \\ Provisioning current progress of the project to all involved parties
        \\ Updating artifacts based on progress and contributions
        \\ Supervising Construction
        \\ Controlling Costs
    } & \colContent{
        Optimising the value of the work the development team performs
        \\ Ordering items in the Product Backlog to best achieve goals and missions; 
        \\ Ensuring that Product Backlog is visible, transparent, and clear to all, and shows what the Scrum Team will work on next
        \\ Updating the Product Backlog at any time
        \\ Helping the Development Team (who make the final estimate) to scope and select trade-offs
    } \\
    
    \midrule
    \headerMiddle{Artifacts} \\
    \midrule
    \colTitle{-4}{Types}    
    & \colContent{
        One requirement documentation
    } & \colContent{
        Drawings
        \\ Cost controlling documents
        \\ Tendering documents
        \\ Time schedule
        \\ Construction diary
    } & \colContent{
        Product Backlog
    } \\\hline
    
    \colTitle{-3}{Lan-\\guages}    
    & \colContent{
        Natural language
        \\ Conceptual models
    } & \colContent{
        Natural language
        \\ Conceptual models
        \\ 2D/ 3D Prototypes
    } & \colContent{
        Natural language
    } \\\hline
    
    \colTitle{-10}{View on the\\artifacts}    
    & \colContent{
        ``Requirements change over the course of the entire development and life cycle of a system" \cite{Poh16}
    } & \colContent{
        The required extent and level of detail depends on type, size and progress of the project under consideration of all discipline-specific requirements \cite{Spr13}
    } & \colContent{
        ``The Product Backlog evolves as the pro-duct and the environment in which it will be used evolves. The Product Backlog is dynamic; it constantly changes to identify what the product needs to be appropriate, competitive, and useful." \cite{WEB_SS17}
    } \\\hline
    
    \colTitle{-10}{Requirement\\types}    
    & \colContent{
        Functional
        \\ Non-functional
        \\ Constraints
    } & \colContent{
        Urban(surroundings)
        \\ Creative
        \\ Functional
        \\ Technical
        \\ Economic
        \\ Ecological
        \\ Social
        \\ Public
    } & \colContent{
        Requirements
    } \\

\bottomrule
\end{tabular}
\end{table*}

\subsection {Liability}

    \textit{Similarities.}
    In connecting their respective customer or client with the developers or other contractors, all three roles or professions act as an interface between the involved parties.

    \textit{Differences.}
    Requirements Engineers and the Product Owners work often in the same organization as the developers. 
    They rarely have a contract directly with the customer.
    Architects, however, sign contracts with the clients directly and represent the client as their trustee.

    \textit{Discussion.}
    As Berry stated, the liability can have an impact on the quality of the product \cite{Berry00}.
    If Product Owners or Requirements Engineers do not see the contracting authority as a client they might skip certain methods, e.g.~testing and prototyping. 
    This might mislead Requirements Engineers to be indifferent towards the product quality or compliance. 
    There might be no incentive to question whether the  customer requirements will lead to a qualitative and satisfying product.
    It should be, however, the Requirements Engineer's job
    to translate customers wishes into beneficial requirements \cite{Berry03}.

\begin{Hypo}
If Requirements Engineers would perceive the contracting authority as their client, this will reduce the flaws of the process and the resulting product.
\end{Hypo}

\subsection {Self-Portrayal}
    \textit{Similarities.}
    Although the literal descriptions in the references differ to some degree, all three roles require similar character traits.
    For instance, trustee (Architects) implies being respectful (Product Owners), showing empathy (Requirements Engineers) being open and committed to the project (Product Owners).

    \textit{Differences.}
    Requirements Engineers present themselves as kind of facilitators.
    Architects see themselves holistically and as trustees of the Client. 
    Holistically means that they are concerned about from what should be built in theory, to how it will be built in practice. 
    The intention is to enable innovative design and construction for the good of the customer \cite{NNK12}.
    Product Owners face a dichotomy.
    On the one hand they are the non-negotiable authority \cite{WEB_SS17}.
    On the other hand they have to respect that nobody can tell the development team how to realize the requirements \cite{WEB_SS17}.
    
    \textit{Discussion.}
    To maintain or evolve a system it could be necessary to try out new or unknown solution spaces.
    This is particularly true if the software development organization works with customers who have a vague or no idea on what they need or if the organization competes on an ever-changing market.
    Requirements Engineers already have techniques and tools to develop ideas with their stakeholders \cite{Poh16}.
    But, if Requirements Engineers were further encouraged to embrace their creative drive, they could also suggest their own visions and support vague customers with an external perspective.

\begin{Hypo}
If Requirements Engineers embrace their creative drive, they will be able to turn more vague ideas into concrete actions and foster the evolution of the customer's system.
\end{Hypo}

    For this creativity to result in actual products, Requirements Engineers should take a holistic approach. 
    They should be able to anticipate what ideas might be difficult to realize. 
    They should make own suggestions for the implementation.  
    This could partially relieve the development team who might have ignored requirements important to the customer or user because they are too difficult, time-consuming, etc.

\begin{Hypo} If Requirements Engineers take a holistic approach by feeling more accountable for the implementation of the requirements, the outcome will better suit customer and user needs.
\end{Hypo}
  
\subsection {Core Activities}       
    \textit{Similarities.}   
    All roles create and manage requirement artifacts.
    They adapt and update requirement artifacts and ensure that all involved parties have access to the latest status.
    Key elements of their duties include maintaining and fostering communication between all involved parties (see Figure \ref{table:results}).
        
    \textit{Differences.}
    The activities of Requirements Engineers are mainly focused on providing a description of what the software should do, independently of implementation details \cite{KS96}.
    This is similar to Product Owners' primary task, taking care of the requirements document, so called ``Product Backlog".
    Architects in their holistic approach not only define what to built but are also tasked with aspects related to the project realization like controlling costs and planning the actual construction.
    
    \textit{Discussion.}
    The Scrum Guide emphasizes the autonomy of the development team.
    This can be risky if the development team does not consider how their decisions impact various stakeholders, or even the overall quality of the product.
    The development team might be under pressure to perform well and may be measured against goals or statistics that are not necessarily connected to customer or user satisfaction.
    Mistrusting customers who seek to avoid these problems may even hire or provide their own Agile Product Owner \cite{EKMV13}.
    A better way to address this might be to present Requirements Engineers as consultants to the development team as well as an integral part of it. 
    Their activities should also include the decision power to interfere as advocates of the customer and users if requirements are not adequately considered. 

\begin{Hypo}
 If Requirements Engineers have decision power in the course of the implementation, then user, customer, and developer satisfaction will increase.
 \end{Hypo}

\subsection {Artifacts}
    \textit{Similarities.}
    All three roles share the understanding that requirement artifacts are changing. 
    They need to be updated as long as the project or product evolves.
    There is consensus that requirements have different stages of depth and detail. 
    
    \textit{Differences.}
    Requirements Engineers document with natural language and conceptual models.
    In addition to that, Architects use prototypes as their main artifact.
    IREB suggests prototypes for ``elicitation" and ``validation" of requirements but not for ``documentation".
    Product Owners only use natural language for their ``Product Backlog".
    Requirements Engineers often target three requirement types: functional, non-functional and constraints. 
    Architects, however, have eight requirement types, including creative, ecological, and social requirements (see Table \ref{table:results}).
    ``The Scrum Guide" does not provide any information on whether or how to differentiate requirement types.
    
    \textit{Discussion.}
    Paetsch et al.~argue that non-functional requirements are prone to get neglected in agile projects \cite{PEM03}. 
    One reason could be that Product Owners only operate with natural languages.
    This aligns with Kotonya et al. who reported that ``more than on e specification language may be needed to represent the requirements adequately"\cite{KS96}.
    In comparison, Requirements Engineers apply natural languages and conceptual models.
    However natural languages ``have inherent ambiguities that can lead to misinterpretations" \cite{KS96} and conceptual models might be too abstract to reflect non-functional requirements.
    Prototypes may likely express non-functional requirements, e.g. usability, better.
   
\begin{Hypo}
 Requirements Engineers who integrate prototypes into their documentation, will represent non-functional aspects better than those who do not.
 \end{Hypo}

Architects have more requirement types guiding their attention to specific and different aspects surrounding a given project.
    This helps Architects to anticipate multiple requirement sources and to acquaint themselves to unknown domains.
    In contrast, the limited requirement types of Requirements Engineers could prevent constraining the mind to specific categories of thinking.
    But in practice this may lead to certain aspects not being considered unless specifically requested by the stakeholders.
    This is not to say that Requirements Engineers should use exactly the same requirement types as the Architect.
    Naturally there needs to be research about the most useful types of systems requirements.

\begin{Hypo}
 If Requirements Engineers expand and specify the requirement types, by e.g. economic, public etc. aspects, they will discover crucial information and correlations in unknown domains more efficiently.
 \end{Hypo}

\section{Conclusion \& Future Work}

This first theoretical comparison provide us with initial discussion ideas regarding liability, self-portrayal, core activities and artifacts for Requirements Engineers. 
    
    In the future, we have to evaluate if these hypotheses can really foster the concept of Requirements Engineering as a distinct profession.
    We also plan to review the overlaps to other roles, e.g. project managers. 
    We are particularly interested in whether the requirement types used by Architects can effectively be applied by Requirements Engineers to enhance their work.

\bibliographystyle{./bibliography/IEEEtran}
\bibliography{./main}
\balance
\end{document}